\documentclass[conference]{IEEEtran}
\IEEEoverridecommandlockouts
\usepackage{cite}
\usepackage{amsmath,amssymb,amsfonts}
\usepackage{algorithmic}
\usepackage{graphicx}
\usepackage{textcomp}
\usepackage{xcolor}
\usepackage{subcaption}

\usepackage{booktabs}
\usepackage{multirow}

\def\BibTeX{{\rm B\kern-.05em{\sc i\kern-.025em b}\kern-.08em
    T\kern-.1667em\lower.7ex\hbox{E}\kern-.125emX}}
\begin{document}

\title{LLM-driven Multimodal Recommendation
\thanks{}
}

\author{
    \IEEEauthorblockN{Yicheng Di}
    
    
    }

\maketitle

\begin{abstract}
As a paradigm that delves into the deep-seated drivers of user behavior, motivation-based recommendation systems have emerged as a prominent research direction in the field of personalized information retrieval. Unlike traditional approaches that primarily rely on surface-level interaction signals, these systems aim to uncover the intrinsic psychological factors that shape users' decision-making processes and content preferences. By modeling motivation, recommender systems can better interpret not only what users choose, but why they make such choices, thereby enhancing both the interpretability and the persuasive power of recommendations. However, existing studies often simplify motivation as a latent variable learned implicitly from behavioral data, which limits their ability to capture the semantic richness inherent in user motivations. In particular, heterogeneous information such as review texts—which often carry explicit motivational cues—remains underexplored in current motivation modeling frameworks. Extensive experiments conducted on three real-world datasets demonstrate the effectiveness of the proposed LMMRec framework. The results show that LMMRec consistently outperforms a range of competitive baselines across multiple evaluation metrics, achieving a relative improvement of 4.98\% in optimal performance. This gain underscores the advantages of integrating large language model-derived semantic priors into multimodal motivation modeling, particularly in enhancing cross-modal alignment and mitigating semantic drift. The findings validate the framework's capacity to capture fine-grained motivational signals and underscore its potential as a model-agnostic solution for recommendation tasks.

\end{abstract}


\section{Introduction}
In recent years, the landscape of recommendation system research has undergone a significant paradigm shift, transitioning from the modeling of surface-level interaction signals toward a deeper exploration of users' latent motivational structures \cite{32yuan2023federated,33yuan2024hetefedrec,37di2025federated}. This evolution is driven by the growing demand for interpretability in AI systems, as understanding the "why" behind user choices is increasingly recognized as essential for building trustworthy and persuasive recommender systems \cite{34yuan2025fellas,38di2025fedrl,35yuan2024hide}. Motivation disentanglement has emerged as a pivotal technical pathway in this context, aiming to decompose the complex interplay of psychological and contextual factors that drive user behavior \cite{36yuan2025ptf,39di2025personalized,40di2025federated}. By isolating these implicit drivers from historical interactions, researchers seek to uncover the cognitive mechanisms that underpin decision-making processes, thereby enhancing both the transparency and the effectiveness of recommendation algorithms \cite{41di2025efficient,42di2025fine,43bao2026personalized,44di2025pifgsr}.

Several pioneering works have laid the groundwork for this line of inquiry. For instance, ComiRec \cite{8cen2020controllable} introduces a multi-interest extraction framework that disentangles user interaction sequences into a set of dynamically evolving motivational vectors. This approach allows the model to capture the multifaceted nature of user preferences, reflecting how interests shift across different temporal and contextual windows. In a complementary vein, DisenGCN \cite{9ma2019disentangled} leverages the power of graph convolutional networks to separate latent semantic factors within node representations. By achieving disentanglement at the representation level, this method enhances the model's capacity to distinguish between distinct motivational dimensions, such as the desire for novelty versus the need for reliability. Collectively, these studies, along with a growing body of subsequent research \cite{47zhang2024artbank,48zhang2024towards,49zhang2023caster,7zuo2024gugen}, have demonstrated that explicit motivation modeling can lead to more robust, interpretable, and adaptable recommendation systems.

Despite these notable advancements, a critical examination of the existing literature reveals a persistent limitation: the overwhelming reliance on structured interaction data \cite{45di2025trustworthy,46di2025global}. Most current models operate within the confines of behavioral sequences, treating motivations as latent variables to be inferred solely from clicks, purchases, or views \cite{59di2023mfpcdr,60fan2024pathmamba,61wang2023federated}. This approach, while effective to a degree, inherently neglects the rich, unstructured semantic information that users naturally generate \cite{62fan2025dipathmamba,63shi2025efficient,64tao2024deeper}. Modalities such as review texts, search queries, and social media posts contain a wealth of explicit and implicit motivational cues \cite{65wang2024tecdr,66song2023msam,67fan2025synchronous,68wang2026reinforcement}. For example, a product review may articulate not only satisfaction but also the specific need that prompted the purchase, such as durability for outdoor use or aesthetic appeal for gifting \cite{69yu2025privrec,70huang2024hierarchy,71wang2023cadcn}. Similarly, a search history can reveal an evolving intent, moving from broad exploration to targeted comparison \cite{72shen2026data,73hong2026stymam,74fan2025foundation}. This linguistic data serves as a direct window into the user's cognitive state, offering a level of detail that interaction logs alone cannot provide \cite{10pan2024less,11zhang2024disentangling,12zhang2025llm,50dang2025data,51dang2025efficient,52dang2024data,53zhang2025mitigating}.

The failure to integrate these heterogeneous data sources results in models with significant semantic blind spots \cite{75zhang2025attention}. When motivation is modeled exclusively from behavior, the system may capture *what* a user does but remains largely ignorant of *why*. This limitation becomes particularly pronounced in complex decision-making scenarios where user intent is nuanced and context-dependent \cite{76di2026srsupm}. The reliance on behavioral data alone leads to a form of semantic sparsity, where the rich tapestry of human motivation is reduced to a sequence of discrete actions. Consequently, the models' ability to generalize across contexts or to provide genuinely insightful recommendations is fundamentally constrained.

This identified gap underscores a pressing challenge for the research community: how to move beyond unimodal behavioral modeling and effectively harness multimodal heterogeneous information. Achieving this requires not merely the concatenation of different data types, but a sophisticated approach to fine-grained motivation disentanglement and cross-modal semantic alignment. The core difficulty lies in establishing correspondences between the structured signals of interaction and the unstructured expressions of natural language, ensuring that the motivational factors inferred from behavior are meaningfully grounded in the semantic content provided by users. It is precisely in response to this complex challenge that we propose LMMRec, a large language model-driven, multimodal recommendation framework. LMMRec is designed to bridge the gap between behavioral and semantic modalities, injecting deep linguistic understanding into the motivation modeling process to achieve a more holistic and accurate representation of user intent.

\section{Methodology}

\subsection{Model Optimization}
Finally, the model is optimized end-to-end through multi-task joint learning, with the overall objective function defined as follows:
\begin{equation}
\mathcal{L}=\mathcal{L}_{MCS}^{\prime}+\gamma\mathcal{L}_{ICM}+\parallel\Phi\parallel_{2}^{2},\
\label{eq10}
\end{equation}
where \(\|\Phi\|_2^2\) denotes the \(\mathrm{L}_2\) regularization term over all trainable parameters in the model.

\section{EXPERIMENTS}

\subsection{Baselines.} 
We compare LMMRec with the following model-agnostic baselines: UIST \cite{26liu2024discrete}, ONCE \cite{27liu2024once}, and AutoGraph \cite{28shan2025automatic}. Representative base models WeightedGCL \cite{29chen2025squeeze} and PolyCF \cite{30qin2025polycf} are selected to evaluate the performance gains of the baselines compared to their respective base models in a model-agnostic manner.

\begin{figure}[tbp]
	\centering
	\begin{subfigure}{0.45\linewidth}
		\centering
		\includegraphics[width=\linewidth]{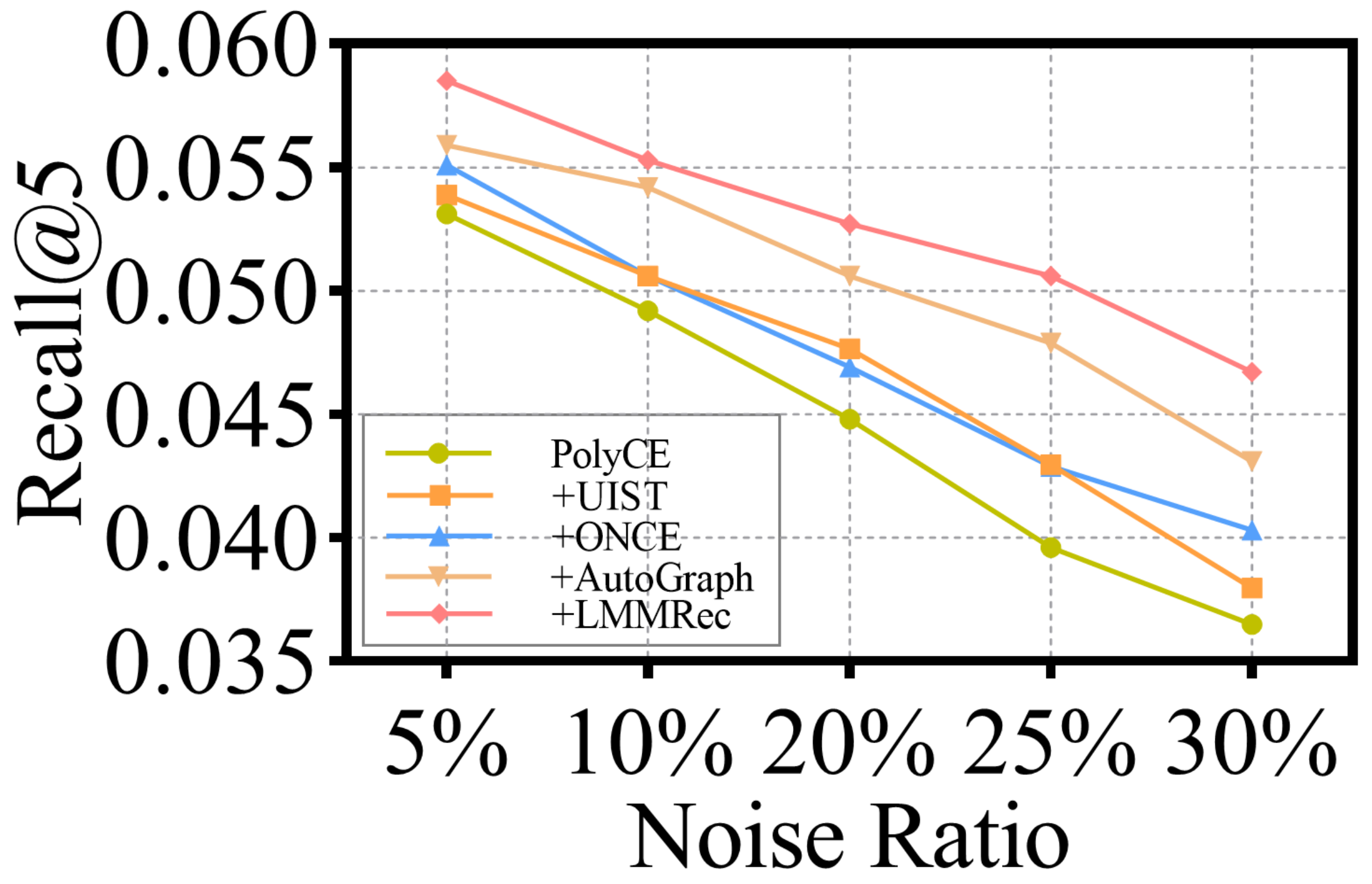}
	\end{subfigure}
	\centering
	\begin{subfigure}{0.45\linewidth}
		\centering
		\includegraphics[width=\linewidth]{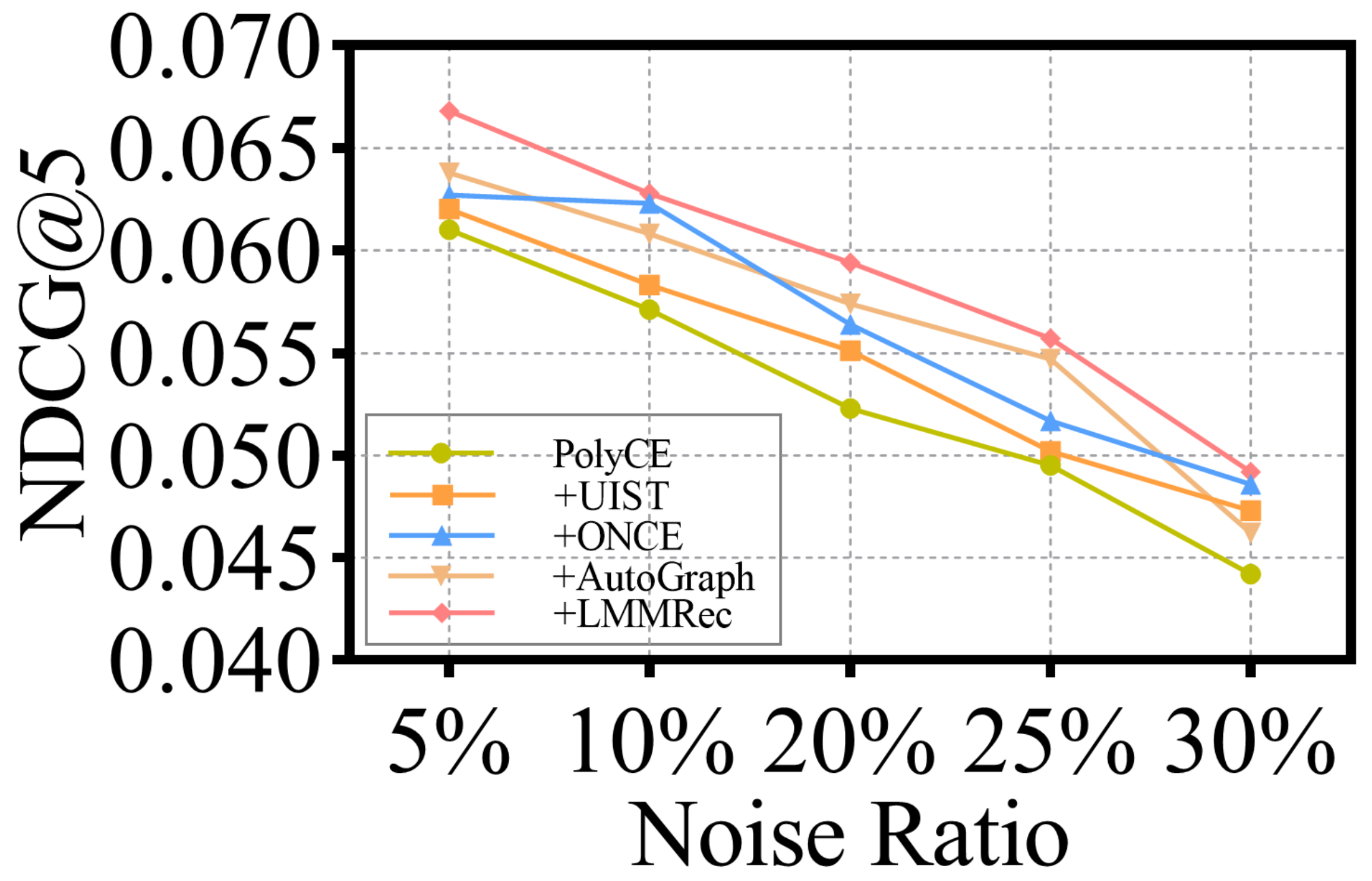}
	\end{subfigure}
       \caption{Comparing performance on the Yelp dataset with varying noise levels using PolyCE as the basic model.}
        \label{fig2}
\end{figure}

\subsection{Performance Comparison}
Specifically, in the two groups of experiments based on WeightedGCL and PolyCF, LMMRec consistently achieves improvements in both Recall and NDCG metrics, with particularly notable gains of up to 4.17\% and 4.98\% on the Yelp and Steam datasets, respectively. In contrast, UIST, ONCE, and AutoGraph improve model performance to some extent but yield limited overall gains. This performance advantage primarily stems from the deep semantic representation capability of the large language model introduced in LMMRec for motivation semantic modeling, enabling it to extract more interpretable and discriminative motivation features from the textual perspective. Meanwhile, the dual-encoder architecture and cross-modal alignment strategy effectively mitigate the semantic gap between text and interaction signals, facilitating consistent modeling in the high-level semantic space.

\subsection{Noise Robustness Analysis}
We evaluate the robustness of LMMRec to data noise by adding nonexistent interactions to the original training data, with noise levels ranging from 5\% to 30\% relative to the training set size. On the Yelp dataset, we compare the performance of the original PolyCE model with those enhanced by UIST, ONCE, AutoGraph, and LMMRec under different noise conditions. As shown in Fig. \ref{fig2}, all model-agnostic enhancement methods exhibit varying degrees of performance degradation as the noise ratio increases, indicating the general impact of noise on interactive recommendation models. However, LMMRec consistently outperforms other methods across all noise levels, demonstrating its superior robustness. This is attributed to the Motivation Coordination Strategy’s consistency constraint in contrastive learning and the Interaction-text Correspondence Method’s mitigation of cross-modal semantic shift, enabling LMMRec to stably capture effective motivation signals under high noise conditions and avoid overfitting to spurious interaction features.

\section{CONCLUSION}
To enhance the modeling capability of recommendation systems for users’ deep behavioral motivations, this paper proposes LLM-driven Multimodal Recommendation. The framework fully leverages the semantic priors and reasoning abilities of LLMs to characterize latent motivation features of users and items from both textual and interaction perspectives, achieving fine-grained modeling at the motivation level. Future work will further explore LLM-based causal motivation modeling and adaptive fusion mechanisms to extend the framework’s applicability in open-domain recommendation and complex interaction scenarios.

\textbf{New content is being updated.}

\bibliographystyle{IEEEtran}
\bibliography{references}

\end{document}